\documentclass[acmsmall]{acmart}
\usepackage{numprint}%
\nprounddigits{5}%
\npdecimalsign{.}%
\npthousandsep{}%
\npfourdigitnosep%
\usepackage{caption}%
\usepackage{subcaption}%
\copyrightyear{2023}%
\acmYear{2023}%
\setcopyright{rightsretained}%
\acmConference[Mindtrek '23]{26th International Academic Mindtrek Conference}{October 3--6, 2023}{Tampere, Finland}%
\acmBooktitle{26th International Academic Mindtrek Conference (Mindtrek '23), October 3--6, 2023, Tampere, Finland}\acmDOI{10.1145/3616961.3616978}
\acmISBN{979-8-4007-0874-9/23/10}%
\begin{document}

\title[Perceptions and Realities of Text-to-Image Generation]{Perceptions and Realities of Text-to-Image Generation}%

\author{Jonas Oppenlaender}
\email{jonas.x1.oppenlander@jyu.fi}
\orcid{0000-0002-2342-1540}
\affiliation{%
  \institution{University of Jyv\"askyl\"a}
  \city{Jyv\"askyl\"a}
  \country{Finland}
  \postcode{40014}
}

\author{Johanna Silvennoinen}
\email{johanna.silvennoinen@jyu.fi}
\orcid{0000-0002-0763-0297}
\affiliation{%
  \institution{University of Jyv\"askyl\"a}
  \city{Jyv\"askyl\"a}
  \country{Finland}
  \postcode{40014}
}

\author{Ville Paananen}
\email{ville.paananen@oulu.fi}
\orcid{0000-0001-7377-9770}
\affiliation{%
  \institution{University of Oulu}
  \city{Oulu}
  \country{Finland}
  \postcode{90570}
}

\author{Aku Visuri}
\email{aku.visuri@oulu.fi}
\orcid{0000-0001-7127-4031}
\affiliation{%
  \institution{University of Oulu}
  \city{Oulu}
  \country{Finland}
  \postcode{90570}
}


\renewcommand{\shortauthors}{Oppenlaender et al.}

\begin{abstract}%
Generative artificial intelligence (AI) is a widely popular technology
that will have a profound impact on society and individuals.
    Less than a decade ago, it was thought that creative work would be among the last to be automated~-- yet today, we see AI encroaching on many creative domains. 
In this paper, we present the findings of a survey study on people's perceptions of text-to-image generation.
We touch on participants' technical understanding of the emerging technology,
their fears and concerns, and thoughts about risks and dangers of text-to-image generation to the individual and society.
We find that while participants were aware of the risks and dangers associated with the technology, 
only few participants considered the technology to be a personal risk.
The risks for others were more easy to recognize for participants.
Artists were particularly seen at risk.
Interestingly, participants who had tried the technology rated its future importance lower than those who had not tried it.
This result shows that many people are still oblivious of the potential personal risks of generative artificial intelligence and the impending societal changes associated with this technology.
\end{abstract}%




\maketitle

\section{Introduction}%
%

Progress in generative artificial intelligence (AI) has exploded in recent years.
    Generative AI refers to a set of technologies that can synthesize text, images, or other media in response to written prompts as input.
This technology has the potential to revolutionize various industries and greatly impact society, particularly in the creative domain.
%
%
%
Less than a decade ago, the general consensus was that knowledge work and creative work would be among the last to be automated \cite{The_Future_of_Employment.pdf,automation.pdf}.
However, recent developments in generative AI have
contradicted these initial predictions~\cite{OECD}.
We increasingly see generative AI being applied in highly creative domains, such as art \cite{prompting-ai-art}, design \cite{paananen2023using}, and research \cite{oppenlaender2023mapping}.
One particularly intriguing domain is text-to-image generation,
as evident in the popularity of generative systems that can synthesize images from short descriptive text prompts.
Such systems include Midjourney\footnote{https://www.midjourney.com}, Stable Diffusion \cite{stablediffusion}, and DALL-E 2 \cite{dalle2}.
    Within a short period of time, Midjourney has become the largest Discord community \cite{mjdiscord}, attracting millions of users.
    StableDiffusion has gained popularity in the open source community since it can be flexibly adapted and personalized to different subject-specific contexts, using 
    fine-tuning on domain-specific images \cite{2208.12242.pdf,2208.01618.pdf}.
Outputs from state-of-the-art diffusion models, such as the above, are often indistinguishable from images created by humans \cite{creativity,prompt-engineering,2304.13023.pdf}.

Generative AI has been trained on digital media collected from the Web without prior consent. Many artists and photographers fear for their livelihood as the uptake of generative AI for commercial use is growing \cite{Revolting,boon}.
    Some call this development ``AI's Jurassic Park moment'' \cite{Marcus.pdf} -- an adapt-or-die moment that could potentially result in massive job loss across many sectors.
    Generative AI can also be an existential risk for organizations and individuals whose business model relies on human effort that can now be automated.
        For instance, book covers can now easily be illustrated with text-to-image systems, without the need to hire or contract a designer or illustrator for this task.
        Question answering websites, such as Stackoverflow, are at risk of becoming obsolete due to people's shifting habits of seeking answers for difficult questions from generative AI.
        Stock photography services are also heavily impacted by generative AI \cite{getty}.
        Even the Tech giant Google is affected as many people shift their search preferences to querying language models 
        instead of tediously sifting through spammy search results \cite{Google,Google2}.
The capabilities of the state-of-the-art text-to-image systems put many organizations and creative professions under pressure.

The arrival of generative AI to creative domains raises a plethora of questions about the transformation of the creative industry, human creative practices, and the future of work.
But while news about generative AI and its potential impact on the workforce is spreading, we should not forget that many people are still oblivious to the powers of state-of-the-art generative AI.
Examining human perceptions of the change brought forth by generative AI sheds light on how this novel phenomenon will affect society. Further, how the role (including risks and possibilities) of generative AI is conceptualized affects the ways it is included in creative practices.
Value-laden reporting in media and literature may decisively influence the adoption and regulation of AI (for good or worse)~\cite{1912.01172.pdf}.

Against this backdrop,
we examine perceptions of text-to-image generation technology, as a popular type of generative AI, among different groups of individuals, including artists, inexperienced users, and (self-reportedly) experienced users of text-to-image generation.
The data was collected at the Researchers' Night, a public event at which researchers present their research to the public. Data was collected via an online survey focusing on people's understanding of the text-to-image generation as an emerging technology, its potential uses, and the dangers of the technology for the individual and society.



\section{Related Work}

The relationship between AI and art has been explored extensively in recent years, in an effort to understand how the perception and attitudes of humans may be influenced by images generated by AI.
In this section, we review several seminal studies in this area.

\citeauthor{1-s2.0-S0747563222003223-main.pdf} examined the aesthetic evaluation of AI-generated haiku poems, distinguishing between those created with human intervention and those made solely by AI \cite{1-s2.0-S0747563222003223-main.pdf}. Their study 
suggested that the most aesthetically pleasing haiku were those made in collaboration between humans and AI, implying a certain synergy that enriches creative output. 
This study also raised questions about the underestimation of AI art, suggesting a phenomenon of `algorithm aversion.'

\citeauthor{1-s2.0-S0747563223000584-main.pdf} identified an anthropocentric bias in art appreciation, positing that recent AI advances in the art domain have challenged traditional human-centric perspectives on creativity \cite{1-s2.0-S0747563223000584-main.pdf}. Their experiments involving over 1,700 participants revealed a pervasive bias against AI-created art, which was seen as less creative and induced less awe, hinting at a persistent human bias towards creativity as an exclusively human trait.
A similar bias was identified by \citeauthor{3334480.3382892.pdf} in their large-scale study involving 565 participants \cite{3334480.3382892.pdf}.
The researchers found that art perceived as human-made was evaluated significantly more favorably than that perceived as AI-made. This highlights the potential existence of a negative perception bias towards AI and a potential preference bias towards human-made creations. 

In a study focused on the younger generation, \citeauthor{ting} explored the perception and acceptance of AI art \cite{ting}. The results showed a high level of acceptance, yet more than half of the respondents could not correctly identify the emotions expressed in AI art. This suggests that while AI is becoming more accepted, there are still gaps in its ability to elicit emotional resonance and comprehension.
In the context of AI-generated images specifically, \citeauthor{2304.13023.pdf} investigated whether these images could deceive human observers \cite{2304.13023.pdf}. Their study found that humans could not significantly distinguish between real and AI-generated images, indicating the sophistication of current AI image generation. However, they also pointed out certain defects in AI-generated images that could potentially serve as cues for discerning authenticity.
Finally, \citeauthor{s42256-021-00417-9.pdf} outlined the potential positive uses of AI-generated media, especially for supporting learning and wellbeing \cite{s42256-021-00417-9.pdf}. They called for the inclusion of traceability measures to maintain trust in generated media, reminding us of the ethical implications of this evolving technology.


The retrospective by \citeauthor{ncw_89.pdf} serves as an important reminder of the continued need for ethical vigilance in the burgeoning field of AI-enabled creativity \cite{ncw_89.pdf}. As AI's role in our creative endeavours continues to evolve, so must our understanding of the ethical boundaries within which it operates. Examining the ethical implications of computer vision in creative applications is important, since this area intersects with everyday life as technology advances. Potential implications encompass issues of privacy, bias, access, representation, and ownership, among others.

The above studies 
underscore the complexity of perceptions towards AI-generated art, including biases, acceptance, and the need for ethical considerations.
Note, however, the great progress that has been realized in recent months. Many studies on the perception of AI art must now be considered outdated, given the strong progress of the field.
Our paper provides a novel empirical perspective on this related work.




\section{Our Approach}%

\subsection{Method}%

We gathered data from visitors of the Researchers' Night 2022 event at the University of Jyv\"askyl\"a. This event is part of the European Researchers' Night\footnote{https://marie-sklodowska-curie-actions.ec.europa.eu/event/2022-european-researchers-night} and intended for researchers to showcase their research to the general public.
We invited visitors to complete an online survey.
The online survey was chosen over other methods (e.g., in-person interviews) for several reasons.
    The Researchers Night is a well-visited public event. At times, the event can feel chaotic and there is a high noise level.
    To be mindful of people's limited time at the event, and to avoid confounding factors interfering with the data collection,
    participants were given fliers inviting them to complete the survey in the comfort of their home. 
Participation was incentivized with a raffle for three Amazon vouchers, each worth 30~EUR.

The questionnaire consisted of 21~items on people's perception of text-to-image generation, emphasizing people's awareness of the risks and dangers associated with this technology
(see Appendix \ref{appendix:questionnaire}).
The questionnaire started with three open-ended items (Appendix \ref{appendix:open-ended}) focusing on people's technical understanding of text-to-image generation, as well as potential future applications and the personal and societal dangers of this technology.
The responses to the three open-ended items were analyzed using in vivo coding \cite{Charmaz}. To this end, the first author read and then iteratively coded all responses. Multiple codes were assigned, if needed, and iteratively improved and merged by frequently visualizing the codes in histogram charts.
Due to the manageable amount of data and the coding being straight-forward, the coding did not require multiple raters and an analysis of inter-rater reliability \cite{McDonald_Reliability_CSCW19.pdf}.

In the second section of the survey questionnaire (see Appendix~\ref{appendix:A2}), participants were presented with a scenario of a person submitting an AI-generated artwork to an art fair.
This scenario was based on a real event \cite{art-fair}.
The third questionnaire section (see Appendix \ref{appendix:A3}) inquired 
about participants' experience with text-to-image generation, followed by the importance of text-to-image generation for participants' current and future professional work (Appendix~\ref{appendix:A4}).
The questionnaire concluded with demographic questions (Appendix~\ref{appendix:A5}).%

Quantitative data were analyzed using an independent two-sample t-test.
The significance level was set at $\alpha = 0.05$, and all tests were two-tailed.
Effect sizes are reported with Cohen's d.

\subsection{Participant Demographics}%
The online survey was completed by 35 participants (P1--P35, aged 19 to 50, $M=33.7$ years, $SD=9.3$ years). 
Participants had diverse educational backgrounds, the most common being computer science, literature, and information systems. Fourteen participants held a Bachelor's degree, 10 held a Master of Science, four a Master of Arts, three a doctoral degree, and one completed no academic degree. Twenty-four participants (69\%) were students.

A third of the participants ($n=12$; 34.3\%) had used text-to-image generation before. The most popular system used by these participants was DALL-E Mini/Craiyon (7 participants), followed by DALL-E 2 (5 participants), Dream/Wombo (3 participants), and Stable Diffusion (2 participants). Participants estimated they had written an average of 20 prompts ($Max=80$ prompts, $SD=22$ prompts).
Participants were, therefore, rather inexperienced with the emerging technology. Participants who had tried text-to-image generation were younger than those who had not tried the technology ($p<0.05$, $d=0.2$).
Ten participants (29\%) considered themselves as being artists. The art created by this group of participants includes paintings, drawings, writing, and digital forms of art. Less common art forms included clothing, music, handicrafts, and food art.


\section{Findings}%
%

The following section describes the findings of our survey and presents results on how the general populace understands the technology behind text-to-image generation, potential application areas, and the perceived importance of this technology. The section continues to present ethical challenges and some of the criticism and concerns towards text-to-image generation.
%
\subsection{Understanding of Text-to-image Generation Technology}%
When asked how the text-to-image system works internally,
the majority of the participants ($n=21$; 60\%) did not have a strong understanding of how text-to-image generation works.
Many of these participants simply stated that the system ``generates'' images in response to keywords.
In the remainder of this section, we focus on the participants who shared their theories of how text-to-image technology works in more detail.

Participants most often related the technology to image retrieval from a \textbf{database}.
    P1 (31y), for instance, likened the technology to \textit{``some kind of huge photo library, each picture has been coded with a word that it describes it the best.
    Maybe some other words connected to it.
    Then it combines the words and finds the best fitting alternatives.''}
\textbf{Search engine retrieval} was a strong theme, mentioned by 13~participants (37\%).
    Participant P10 (25y), for instance, thought \textit{``they use Google or other search engines and combine some of the best results in some way.''}
    One-the-fly retrieval from Google or some other repository or database was mentioned often among these participants.
    P3 (25y), for instance, mentioned that it \textit{``fetches image from the repository and merges two or more pictures, pre defined ideas and develops a new one,''} and P16 (24y) thought \textit{``when the system receives a text prompt, it goes through a large set of images that correspond to the particular prompt (kind of like Google Images I think), analyses them, and creates a new images based on these existing images.''}
P25 (19y) stated that \textit{``it searches the web for all material containing and/or even mildly resembling the prompt(s) given. Then the system analyses all of the gathered material, combines elements from several (if not all) of them to generate the final image.''}
P7 (27y) intermixed two opposing theories on the inner working of the technology, stating that \textit{``either the engine searches your input words from the internet and uses the images found as a reference to compile a completely new image OR the engine has been fed image data from the internet and it uses what it has learned to compile the new image.''}


The theme of \textbf{mixing or combining} existing images was raised by several participants ($n = 11$; 31.4\%).
Connecting to the theme of image retrieval from a database, P4 (31y; artist) thought that \textit{``[it] fetches images from the repository and merges two or more pictures, predefined ideas and develops a new one,''} and P35 (33y; artist) mentioned that \textit{``it tries to find corresponding pictures for the words in the text and then combines them to create a final picture.''}
The text-to-image system was thought to merge and fuse images, even though participants could not explain the inner workings in more detail.
P11 (42y), for instance, thought \textit{``it somehow can merge, fuse parts of the images to construct an image based on the keywords in the input sentence.''}
A different mental image was held by P29 (44y) who thought that the generative technology would replace parts of images:
    \textit{``it depends on a huge database of labelled images with descriptions of the items shown in them and relations between them (the vase is on the table). Then it's a matter of replacing items. If the user writes: "a cat is on the table", and ML algorithm will replace what it knows as "the vase" with a foto of a cat in the place of the vase. The bigger and more diverse the dataset, the better the results.''}

A minority of participants ($n = 8$; 22.9\%) had some understanding of how the technology works. P23 (35y), for instance, wrote that
    \textit{``The AI has learned to produce pictures while the prompts function as parameters for the algorithm. The AI probably has a large quantity of existing pictures as a learning material that has been coupled with keywords. Through feedback the AI has gradually become better at producing pictures that match the prompts.''}
However, in their description of the technology, only four participants explicitly distinguished between training and inference time.
    The `training time' is when the AI learns from vast datasets, while the `inference time' is when the AI applies this learned information to generate new outputs.
Distinguishing between training and inference is an important step in understanding how generative AI works.
Most participants did not make this important distinction.

\subsection{Potential Application Areas of Text-to-image Generation}%
\begin{figure}[htb]%
  \centering%
  \includegraphics[width=.6\textwidth]{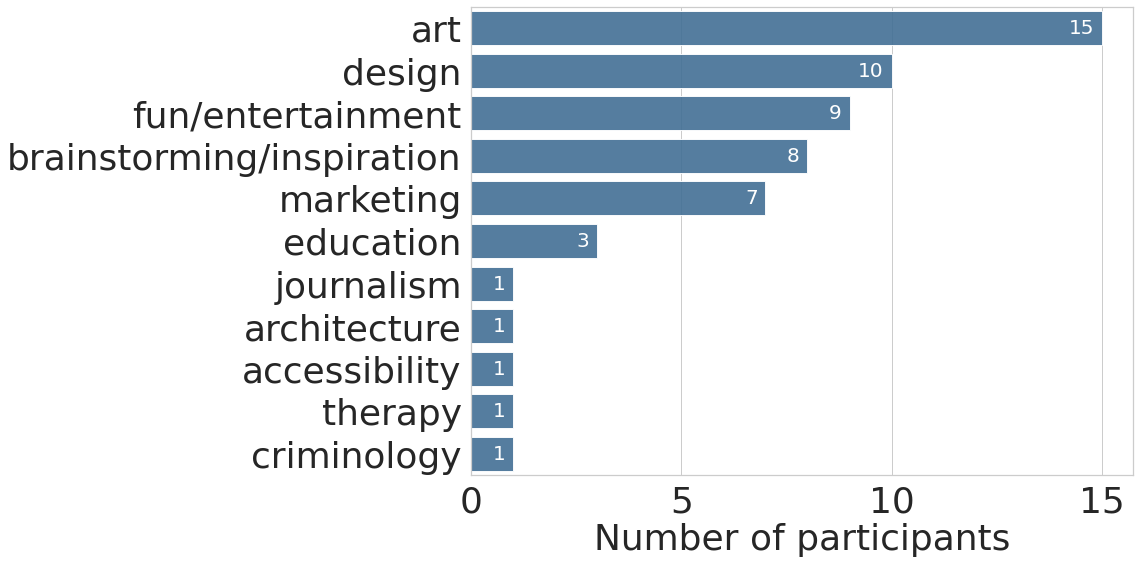}
  \caption{Participants' thoughts on potential application areas of text-to-image generation.}%
 \label{fig:application}%
\end{figure}

Creative domains were most common when it comes to participants' thoughts about potential applications for text-to-image generation (see \autoref{fig:application}).
Creating \textbf{digital art} was the strongest application area seen by participants.
Participants thought the technology was well suited for creating digital artworks, illustrations, logos, and other visual media.
Besides directly creating artworks with the technology, one participant also mentioned that generative AI could 
\textit{``be a good tool for artists to have multiple different references while making their own artwork''} (P15; 21y; artist).
P3 (25y) acknowledged commercial use of text-to-image generation and mentioned the synergy of this technology with other technologies, such as non-fungible tokens (NFT), for selling and buying digital artworks.
%
%
While some participants thought of text-to-image generation as a powerful co-creative tool in the toolbox of artists (e.g., P10, P15, and P23),
many participants thought the technology was a potential replacement for artists and designers altogether. P11 (42y) imagined \textit{``a system where you give it the requirements of a design, you press a button, and you get a multitude of designs to choose from.''}
This system grants non-experts the creative capabilities that were once exclusive to professional designers.
P9 (30y) acknowledged that text-to-image technology, therefore, \textit{``
lowers the barrier to creating images.''}
Text-to-image generation systems \textit{``may provide a method for quickly creating the needed pictures when required for any given project and decrease the effort and skills needed for their creation''} (P9; 30y).
Generative design systems may provide users with an end-to-end way of producing artworks, without having to turn to artists and designers as middlemen in the creation process.
P14 (44y; artist) pondered that \textit{``it will remove the human in the production of illustrations for all sorts of purposes.''}

Further application areas mentioned by participants included \textbf{brainstorming and ideation}, in application areas such as new product development (NPD) and design.
Text-to-image generation could help to visualize ideas and designs, 
\textit{``to see what it could look like''} (P2; 24y).
P27 (39y) acknowledged that \textit{``these systems could lessen the need to create visual material from scratch.''}
%
Generative AI provides a means to synthesize
\textit{``the optimal photo or image for some [specific] purpose''} (P19; 50y).
In general, text-to-image generation 
\textit{``makes it easier and faster to create pictures for ads, maybe animated TV shows, etc.''} (P2; 24y).
Text-to-image generation was seen as a fast and cheap alternative to 
\textit{``manipulating photos or doing digital art''} (P7; 27y).
Text-to-image generation \textit{``decreases the costs of creating pictures traditionally (aka. with a camera, studio set-ups, the cost of artists' work, etc.)''} (P9; 30y).
Text-to-image generation considerably lowers the price for illustrations and \textit{``producing pictures for advertising products''} (P5; 29y).
Therefore, one large application area was seen in \textbf{advertising and marketing}, to generate \textit{``cheap images for ads and illustrations. It might replace stock photography websites and companies''} (P4; 31y; artist), and would be suitable for ``game concept art or marketing art'' (P15; 21y; artist).
Participants further mentioned that the technology could be used in journalism and media, to generate images for magazine articles and websites, and a broad range of visual illustrative media, such as \textit{``illustrations for cards, childrens' books or almost anything''} (P33; 45y; artist). 


Another strong application area was \textbf{fun and entertainment}.
Text-to-image generation makes a fun pass-time, according to participants.
The entertainment industry was seen as an application area, for instance for making animations and games \textit{``only with a script''} (P8; 25y).
Some participants likened the fun derived from text-to-image generation to meme creation (e.g., P18; 36y).
P20 (42y) pondered about social applications, and thought it \textit{``would be fun to `play' with it with other people and create a social-pictures, or something like that.''}
P24 (37y) mentioned that \textit{``the systems are great fun and humorous,''} but had concerns that \textit{``if the systems get `better' in the future, it ruins that fun.''}

In the remainder of this section, we highlight some less commonly mentioned application areas, including \textbf{education, therapy, journalism, criminology, and accessibility}.
P31 (23y; artist) recognized the broad potential of text-to-image generation, and stated that \textit{``the potential of these systems is infinite. They could be used in schools to aid teaching, in therapy, to speed up design processes such as games, etc.''}
%
As for applications in education, 
P24 (37y) thought that the technology was useful because \textit{``pictures are in many cases a more effective way to describe things than words.''}
Text-to-image generation was seen both as a tool to aid teaching in schools (P31; 23y; artist) as well as to \textit{``inspiring kids, giving ideas''} (P1; 31y).
In the educational context, text-to-image generation could be applied to illustrate educational materials.
%
%
One participant mentioned that text-to-image generation could be useful in criminology \textit{``to reconstruct crime scenes in some cases''} (P33; 45y; artist).
P14 (44y; artist) also alluded to a forensic use by mentioning that \textit{``it can be used to detect connections between images. It can link images e.g. an image of a person with some rash can be linked to some disease.''}
Last, participants mentioned applications for accessibility, as a kind of universal tool that could help people with accessibility needs:
\textit{``It could be used as a kind of visual dictionary or translator; you say a word and the machine draws it''} (P34; 25y; artist).

\subsection{Perceived Importance of Text-to-Image Technology}%
\begin{figure}[!htb]
     \centering
     \includegraphics[width=.49\textwidth]{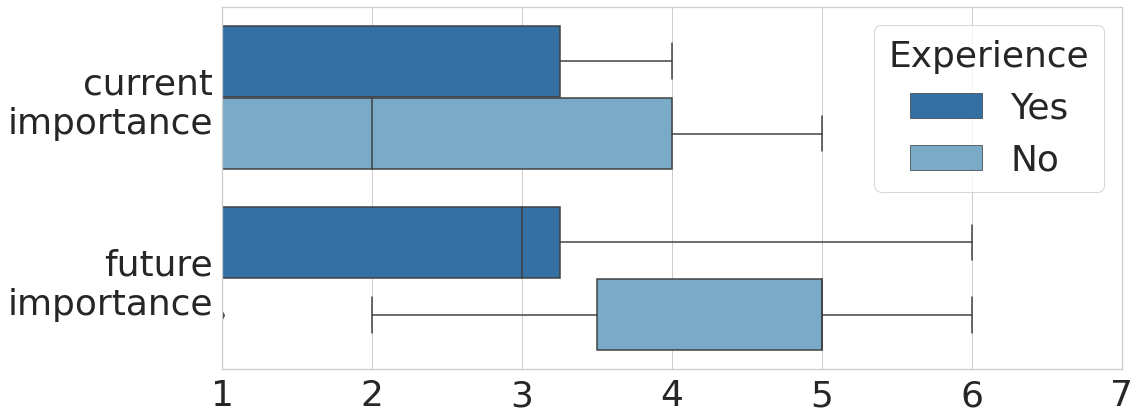}
     \hfill
     \includegraphics[width=.49\textwidth]{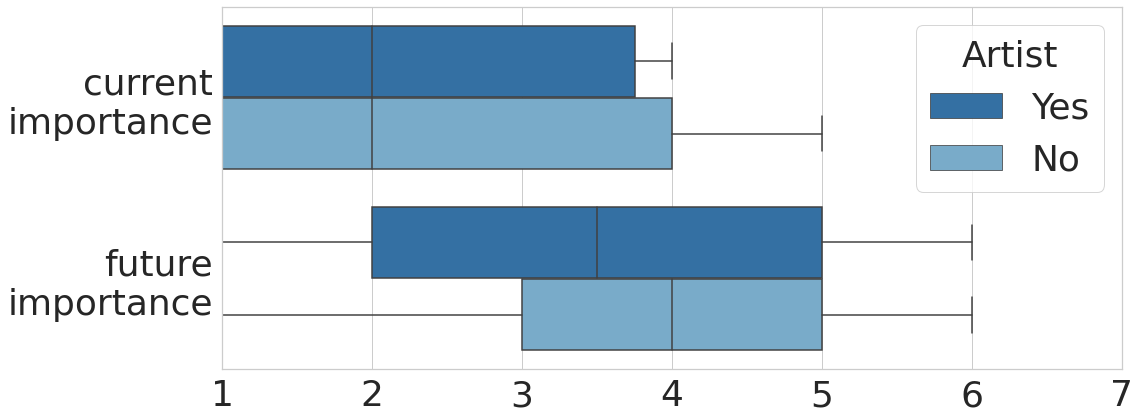}
     \caption{Boxplot comparison of current and future professional importance of image generation for participants who did and did not try text-to-image generation before (left) and self-identified artists and non-artists (right), on a Likert scale from 1 -- Not At All Important to 7 -- Extremely Important.}
     \label{fig:importance}
\end{figure}

\label{fig:importancedangers}%
Most participants responded that text-to-image generation did not hold any importance in their personal and professional lives, but acknowledged that it could play an increasingly important role in the future (see \autoref{fig:importance}).
Interestingly, those who had tried it before found text-to-image generation not as important for their professional future, as opposed to those who had not tried it before. This difference was significant ($p<0.05$, $d=0.53$) and not found among self-declared artists.

\subsection{Ethics of Disclosing AI Generation}%
\begin{figure}[!htb]%
     \centering%
     \includegraphics[width=.55\textwidth]{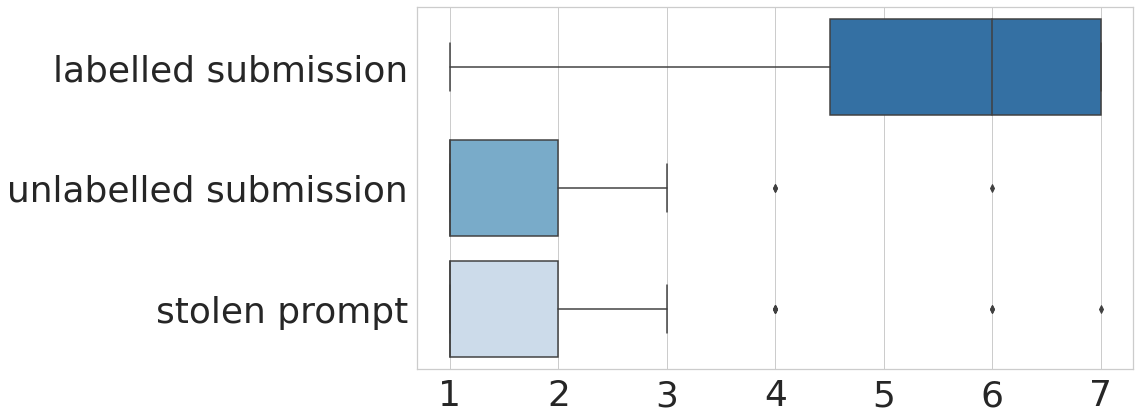}
     \caption{Rating of the ethics of submitting a text-to-image artwork to an art fair on a Likert scale from 1 -- Not Ethical At All to 7 -- Highly Ethical.}
     \label{fig:disclosure}
\end{figure}

About half of the participants ($n = 19$; 54.3\%) were of the opinion that it should be disclosed when something was created with AI.
    Ten participants (28.6\%) had no strong opinion about this, and six participants (17.1\%) thought that AI-generated images do not need to be labeled as such.
When presented with the scenario of an AI-generated artwork being submitted to an art fair, participants thought that it was unethical to submit without disclosing that the image was created with AI (see \autoref{fig:disclosure}).
Interestingly, not labeling a submission to an artwork contest as ``created by AI'' was seen just as unethical as submitting an artwork created with a stolen prompt.

\subsection{Criticism and Concerns about Text-to-image Generation}%
\begin{figure}[!htb]
\centering
\includegraphics[width=.6\textwidth]{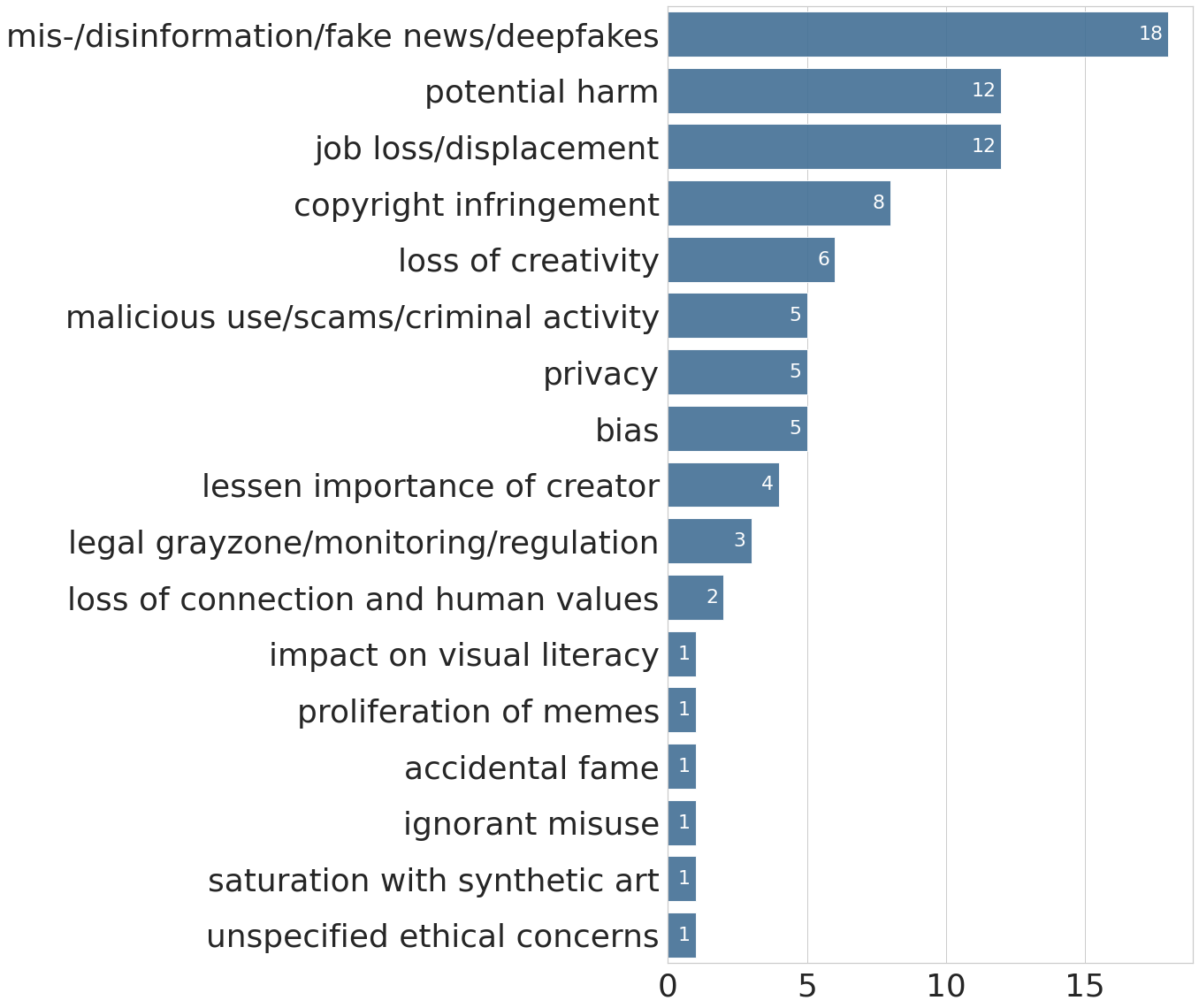}%
    \caption{%
Participants' thoughts on risks and dangers of text-to-image generation.
    }%
\label{fig:dangers}%
\end{figure}

%
Many participants did not see a risk or danger for themselves.
P24 (37y), for instance, mentioned \textit{``I couldn’t think of anything that could be dangerous for myself.''}
But while the majority of participants did not think that text-to-image poses a personal danger to themselves, participants still voiced many concerns about the effect of this emerging technology on society as a whole (see \autoref{fig:dangers}).
It was a common theme to not see dangers personally, but note them for society.  P17 (23y) remarked, for instance, \textit{``I don't see much danger to myself. I find most internet-based activity affecting negatively the society.''}

The use of AI-generated imagery for opinion manipulation, fake news, and ``deep fakes'' was leading cause for concern (see \autoref{fig:dangers}).
Many participants warned that synthetic images could be spread naively (\textbf{misinformation}) or for malicious purposes (\textbf{disinformation}).
Text-to-image generation could be used for ``creating false re-creations or look-a-like versions that can cause harm'' (P19; 50y).
In the hands of authoritarian leaders, the technology was seen as especially dangerous.
Text-to-image generation \textit{``could increase the amount of disinformation circulating on social media. Certain deepfake-systems have already caused havoc for example in the politic fields, so an AI system like this, if powerful enough, poses a huge threat''} (P25; 19y).
The realism of images synthesized by generative systems was seen as problematic in the context of disinformation and fake news.
P6 noted that \textit{``it makes so authentic-looking pictures, that it might be difficult to differentiate generated fake pictures from genuine ones. These pictures might be used for propaganda for example in fake news.''}
According to P7 (27y), \textit{``more and more fake images will start circulating''} which will cause an \textit{``ethical problem:
    Is it okay to make AI generated pictures of other people? Is there a difference between making these images of private people vs. public figures? It becomes increasingly harder to distinguish what media is factual and what AI generated.''}
This would make it 
\textit{``harder to know in the future what information is reliable - a photo isn't as reliable proof as it used to be''} (P11; 42y).
P12 (48y) worried about this development, noting \textit{``that more and more things can be faked more and more easily -- what is the concrete realism that we can trust any more in the longer run? Are we anchored less and less to reality as the time goes on?''}
P13 (45y) also mentioned this concern: as more and more fake imagery circulates online, \textit{``there might be a chance that certain people would perceive reality `in a wrong way'.''}

Some participants mentioned the potential of text-to-image generation to cause \textbf{harm}, such as depression and other illnesses related to mental health.
The AI could be intentionally used to produce offensive and abusive images for cyberbullying, or inappropriate images that are \textit{``not sensitive to people beliefs''} (P3; 25y).
But the harm could also be accidental, such as the negative effects of accidental fame and leakage of private information. 

Another major concern was \textbf{job loss} and unemployment due to increases in productivity and \textbf{job displacement}.
   Participants mentioned that generative AI is cheaper and faster, and this could lead companies to not commission works from humans. 
%
Artists were seen as especially vulnerable to generative AI.
P1 (31y) remarked that \textit{``if I were an artist, I'd be worried that might put me out of business.''}
Generative systems were seen as \textit{``a threat to artists. When these systems get more popular, different corporations etc. might stop commissioning art from artists and use these systems instead, since they are much faster''} (P16; 24y).
P34 (25; artist) mentioned the fear \textit{``that people will start using AI-generated images as a cheaper alternative when otherwise they would have to commission an artist, and probably even using an artist's works as reference without permission.''}
A potential \textit{``loss of creativity''} could be the result, \textit{``as graphic designers and artists become less needed''} (P26; 30y; artist).
This would affect \textit{``the income and status of artists who already suffer from poor income and low appreciation''} (P22; 33y).
The self-declared artists in our sample were personally worried about future changes to their profession and practices.
P34 (25y; artist), for instance, was concerned about creations ending up in the training data without permission, and that \textit{``eventually people will not be able to recognize the human-made artworks from the AI-generated ones. I have seen people accusing artist for using an AI and lying about their art, simply because the people have thought the artist's style looks like it's `generated'. This could lead to people forcing artists to constantly offer proof that their art is truly made by them.''}

Related to the potential loss of jobs, many participants noted that text-to-image generation operates in a legal gray zone with \textbf{copyright infringement} being one major concern, in two regards.
First, images are being used in the training data without prior consent.
Second, text-to-image generation allows to synthesize images in the style of certain artists without their consent.
    P15 (21y; artist), for instance, mentioned that some \textit{``people have directly used the artists name in the prompt to get an image to resemble the artists work as much as possible [without consent].''}
Ease-of-use and availability was seen as a facilitator of copyright infringement, as \textit{``it can make copying real artists' style/work too easy and lead to copyright issues''} (P28; 23y; artist).

As potential long-term effect, some participants mentioned there could be a loss of appreciation for artists and their work. 
Synthetic images could \textit{``lessen the importance of the creator and the creative act''} (P4; 31y; artist).
Text-to-image generation could lead to a depreciation of the value of art and the human virtues encoded in such art.
The widespread use of text-to-image generation could \textit{``lead to people valuing artists less''} (P16; 24y).
Several participants thought there could be a \textit{``decline in human creativity''} (P2, P23, P27).
    The technology \textit{``could curtain artist imagination, when an AI can create art better than humans''} (P3; 25y),
with potential \textit{``knock-on effects on unemployment, depression, caused by this lack of connection with human values and needs to create and be creative''} (P26; 30y; artist).

Homogenization of styles and values was seen as another problem by some participants.
P21 (49y) mentioned that \textit{``a great danger of enforcing certain values through what kind of imagery AI is prone to create and what to leave out -- e.g. what kind of human bodies appear in the images? If existing art is used as teaching material from art institutions with readily available open data -- that will probably lead to mainly white European bodies with certain aesthetics.''}
The AI was thought to be \textit{``ultimately limited in its aesthetics''} (P20; 42y). This low diversity in synthetic imagery could contribute to \textit{``narrow the viewpoint of the world''} if \textit{``a lot of images start to look the same''} (P15; 21y; artist). 
Synthetic images could lead to a \textit{``biased and one-sided visual culture''} (P23; 35y).
As P20 (42y) noted, generative AI \textit{``has a great danger of enforcing certain values''} by showing \textit{``mainly white European bodies with certain aesthetics''}
(P15; 21y; artist).
P15 also stated that  \textit{``AI images can also have a narrow viewpoint of the world and people. A lot of images start to look the same and a lot of people generated by AI resemble each other a lot. You don't see as many different faces and body types as you see in photography or art.''}
The low diversity in synthetic imagery was seen to have a potentially negative impact on visual literacy if \textit{``school book visualizations are [made] in the future with AI''} (P20; 42y).
P20 voiced some concerns about this in the context of illustrating educational materials, as \textit{``cheaper AI based art is ultimately limited in its aesthetics. If e.g. school book visualizations are in the future made like this, then it might lead to issues with visual literacy.''}
P26 (30y; artist) noted that generative AI could be \textit{``a movement away from the things that make us human, e.g, human emotions being reflected in human-made art.''}
As P26 put it, text-to-image generation is \textit{``a movement away from the things that make us human, e.g, human emotions being reflected in human-made art''}.
As dramatically phrased by P12 (48y):
    \textit{``Real artists bite off bat's head or in this context, bite off ears -- how can AI ever going to do that? And when it does, it's going to be frigging scary.''}

\subsection{Limitation: Reflection on the Survey Study}


This work gathered and synthesized diverse perspectives on the emerging technology of text-to-image generation. However, we find some limitations in our survey study that are worth mentioning.

\begin{figure}[!htb]
     \centering
    \includegraphics[width=.85\textwidth]{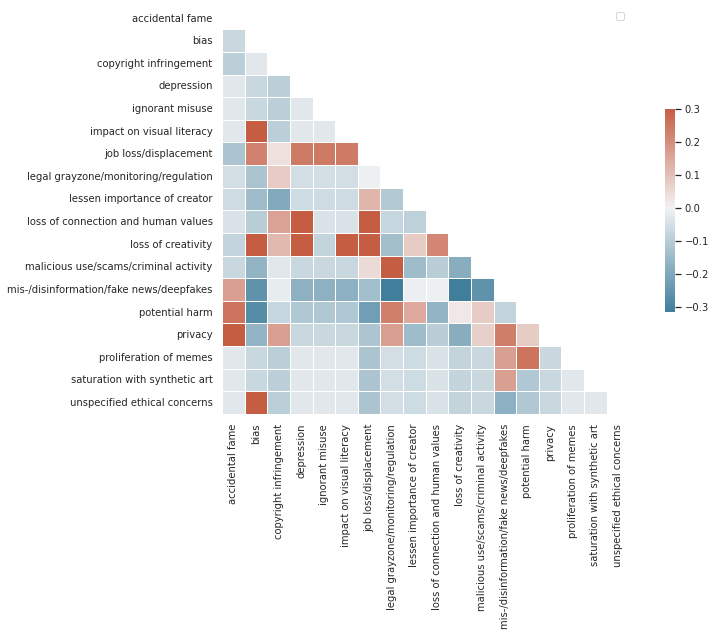}%
    \caption{Correlation between concerns voiced about text-to-image generation by study participants.
    For instance, participants who worried about malicious use of text-to-image generation also often mentioned that it operates in a legal grayzone (indicated with red shading), but were less likely to mention loss of creativity (indicated with blue shading).
    }%
     \label{fig:crosstab}%
\end{figure}

We find evidence that participants provided one-sided responses (see \autoref{fig:crosstab}). Some responses from participants indicate potential biased views about the emerging technology.
From \autoref{fig:crosstab}, it becomes apparent that certain topics (such as misinformation, but also job loss) were less likely to be mentioned together with other topics by participants.
The potential bias in the responses, thus, presents a potential threat to the validity of our findings.
It seems that many participants got hung up on one idea and forgot to reflect on other topics, thereby failing to provide a well-rounded perspective.
Future work could involve more nuanced techniques, such as workshops and in-depth interviews, to overcome the survey's inherent limitations in sparking nuanced thought and encouraging in-depth exploration of the topic.




\section{Discussion and Future Work}%

In this section, we discuss the noteworthy findings of our study: the public's perception of the expectations of the technology and its societal impact, the technical proficiency required and how this can lead to misconceptions, and lastly, how generative AI intersects with human creativity.

\subsection{Reevaluating Expectations: The Public Perception and Anticipated Dangers of Text-to-Image Generation}

Our study illuminated an interesting dichotomy among the participants. The majority of participants did not readily recognize potential application areas of generative technology in their professional work and personal life.
This surprising response seemed to contradict the inherent utility of text-to-image generation in various areas, such as producing graphics for PowerPoint presentations or birthday greeting cards.
One possible explanation lies in the public's natural inclination towards tangible outcomes rather than the processes leading to these outcomes, as witnessed by the cultural focus on images and symbols~\cite{baudrillard1994simulacra}. In the case of text-to-image generation, the deep learning process remains largely invisible, while the produced digital image takes center stage.
This could lead to a skewed perception of the technology's applicability.
This reasoning might also explain why participants commonly envisaged artists as the group most impacted by text-to-image generation. The art sector, being highly visible and often requiring extensive manual effort, stands out as a tangible field where this technology could produce clear changes.
Participants recognized artists as a vulnerable group that could face job loss or job displacement due to text-to-image generation technology.

While seemingly neglecting to deeply consider the technology's impact on their own lives, participants nevertheless demonstrated awareness of the potential dangers and risks of generative AI, such as text-to-image generation, for society as a whole. Participants cited issues including bias, potential loss of human values and creativity, and the potential misuse of the technology for nefarious purposes, such as deep fakes, propaganda, and disinformation.
Interestingly, we observed a divergence in attitudes towards the technology among those who had tried it and those who had not. Those who had personally experienced text-to-image generation perceived its future importance as being lower than those who had not tried it. This phenomenon seems to align with the general hype cycle of technology, in which emerging technologies often encounter a trough of disillusionment before their full potential is acknowledged and realized \cite{HYPECYCLE}.
The divergence in participants' responses could indicate a gap between the anticipation of what the technology can deliver and the current reality of its capabilities. It invites future work on the current perceptions of generative AI and how perceptions might evolve as the technology matures and becomes more pervasive in various facets of life.
Our study makes the first step in this regard, illustrating a complex landscape of public perception and anticipated impacts of text-to-image generation. 

\subsection{The Societal and Individual Impact of Generative Technology}

As the prevalence of deep fakes increases and visual literacy diminishes, the legibility and authenticity of the online sphere becomes increasingly elusive. This ongoing shift, exacerbated by the insidious combination of these factors, could provoke an atmosphere of confusion and mistrust, casting a shadow on the possibilities presented by generative technologies.

Notably, this disconcerting dynamic echoes an aspect that can be considered a downside of technology-driven automation: the gradual fading of traditional human skilled professions. For instance, the fine art of manually creating artworks, such as paintings and drawings, is facing a similar fate as it gets eroded in the face of automated art generation. This gradual loss of skills finds a historical parallel in professions like shoe-making, which has dwindled over time to be practiced only by a select few artisans. This evokes poignant reflections on the way societal progress and technological advancements are entwined with the potential disappearance of age-old crafts and skill sets.

Further deepening the complexity of the impact is the 
potential job displacement and job loss triggered by generative technologies, as revealed by study participants. A clear pattern emerges in this context: while artists foresee a personal impact due to the rise of generative technology, non-artists perceive it as a distant development. This discrepancy draws attention to the concept, astutely captured by William Gibson's quote, ``The future has arrived -- it’s just not evenly distributed yet'' \cite{gibson}.
This observation raises a compelling question: why is the adoption and expectation of generative technology not evenly distributed? The disparity in perception between artists and non-artists alludes to the fact that the implications of this technology are perceived more directly by those whose work it stands to disrupt. It suggests the presence of an intriguing dynamic between societal sectors and their proximity to technological disruption. As a potential avenue for future exploration, it would be worthwhile to investigate how generative technology is being adopted across different sections of society, and the factors influencing the uneven distribution of its adoption and implications.

\subsection{The Dichotomy of Technical Proficiency in Text-to-Image Generation}

We speculate that there could be a growing divide.
On the one side are technologically adept engineers who understand machine learning and can build powerful systems with this technology.
On the other side, there are those who use the technology, but may not fully understand how it works. The latter group has grown significantly in the past several years with the spread of generative technology.
It's worth noting that the creation of technology that reduces the need for human labor can be seen as a position of power and privilege. Those who build these systems effectively shape how we use technology in our daily lives. Interestingly, one of the reasons why text-to-image technology has gained popularity is its ease of use. Natural language is an intuitive tool to control the technology.
But this raises an interesting question: Do we need to fully understand this technology to use it effectively? Our research suggests that to some extent, yes, understanding is important. Knowing about the possible issues, like copyright or intellectual property concerns, is critical when using these systems.
%
%
We believe a basic level of technical comprehension is essential for  understanding the technology's broader implications, such as legal ramifications involving copyright issues and intellectual property rights.
However, this should not be perceived as an insurmountable hurdle.
As adoption rates rise, public familiarity with the technology will naturally increase, and enable a more informed use of text-to-image generation technologies, ensuring a more equitable and responsible technological landscape.%

\subsection{Misconceptions in the Understanding of Text-to-Image Generation Technology}%

As generative AI technologies continue to evolve and permeate various aspects of society, it is crucial to shed light on the public's understanding of such systems.
Despite its growing prevalence and influence, our findings suggest that deep technical understanding of text-to-image generation technology is relatively rare. Participants' conceptions often skewed towards visualizing the technology as merely retrieving images from an existing database or a search engine.
This misinterpretation is alarming, considering that these systems don't simply retrieve, but rather generate novel images using extensive training on diverse image datasets.
A concerning finding was the common association of this technology with image retrieval from a pre-existing database, a stark contrast to the actual process of generating unique images based on extensive training on diverse image sets. This misconception underscores the need for clearer differentiation between training and inference times in public communication, as understanding these two phases is essential to grasp the inner workings of generative AI.

To comprehend the nuances of the technology and its potential societal impact, a solid technical grasp of how generative AI is trained is crucial. Unfortunately, our study indicates that this level of understanding is not widespread, necessitating an increase in both educational initiatives and public awareness efforts.
The implications of these misconceptions extend beyond the technical realm. Misunderstandings about generative AI can influence news coverage and even the formation of potential regulations, stressing the importance of a sound technical grasp of this technology in the public sphere.
Future research should, therefore, focus on devising strategies to make these systems more understandable to the public.
Additionally, it would be insightful to investigate if participants' perspectives would shift with greater insight into the effort and complex processes involved in producing generative images.
This will enable a broader understanding of generative technology, which will be critical in guiding its responsible and beneficial development and use in the future.




%
%
\subsection{The Intersection of Generative AI and Human Creativity in Image Generation}%
Generative AI has been steadily revolutionizing various fields, one of the most prominent being image generation. As an integration of diverse models and intricate processes, image generation culminates in the production of a singular artifact -- the image itself. The advanced nature of these technologies might be considered akin to magic by some observers \cite{money}, recalling Arieti's notion of ``magic synthesis,'' a term he coined as a metaphor for creativity \cite{arieti}.

As our survey study demonstrated, a key element of the ongoing discourse surrounding generative AI pertains to the tension between technology and traditional creativity. For instance, while synthetic images can reproduce the appearance of traditional artwork, they often lack the characteristic individuality and minute nuances that stem from manual creation. An oil painting carries the imprint of the artist's brush strokes, revealing subtle movements of the hand and minute imperfections. As society leans towards adopting generative AI more broadly for tasks such as text-to-image generation, there is a legitimate concern that both the appreciation and the expression of such intricate detail could erode. 
A parallel can be drawn between this phenomenon and manual labor intensive professions, such as engineering. Car repair serves as an example. While many individuals can operate a car, contemporary vehicles often require specialized skills and tools for repair and maintenance. Over time, an understanding of the underlying mechanics has become lost and, today, it is difficult to repair a modern car without special skills and knowledge. 

A poignant critique by \citeauthor{Pallasmaa} lends further insight into the potential downsides of excessive computerization in creative fields \cite{Pallasmaa}. The authors express concern about an increasing sense of detachment in design processes due to uncritical use of computers. This detachment risks severing the link between design and its innate connection to the human psyche and body~-- a connection often facilitated through the tangible act of drawing and imaginative empathy.
The existential and authentic nature of art and architecture conveys what it means to be human in the world, a sentiment difficult to encapsulate within the confines of a mechanized process, however delicate and subtle it may be. In the realm of architecture, \citeauthor{Pallasmaa} note that computerized renderings often reduce human figures to mere adornments, akin to flowers in a vase, thereby underscoring the risk of loss of authenticity and depth in the face of mechanization.
These observations and concerns bring to the fore an important question: how can image generation, or more broadly, generative AI, be leveraged in a manner that preserves and enhances human creativity rather than overshadowing it? This question becomes pivotal as we venture further into the integration of AI and human creativity. The challenge, it appears, lies not in the technology itself but in its application - in balancing \citeauthor{arieti}'s magic of synthesis with the authenticity of human expression.

\section{Conclusion}%


The rapidly evolving field of generative AI, believed just a few years ago to be incapable of penetrating creative spheres, is today making significant advances into many artistic domains.
This paper presented insights obtained from a survey study focused on people's perceptions of text-to-image generation technology.
The study explored the participants' understanding of this emerging technology, their apprehensions, and their perspective on the potential risks and dangers posed to both individuals and society. It was observed that while most respondents were cognizant of the technology's broader risks, they rarely perceived these as personal threats. The perceived risk was greater for others, particularly for artists. Intriguingly, participants who had experienced the technology firsthand deemed its future relevance to be less than those who hadn't tried it. This finding underscores a widespread unawareness of the personal risks linked with generative AI and the impending societal transformations associated with it.

\bibliographystyle{ACM-Reference-Format}%
\bibliography{paper}%
\appendix%
{%
\small%
\section{Questionnaire}%
\label{appendix:questionnaire}%
\subsection{Open-ended questions:}%
\label{appendix:open-ended}%
\label{appendix:A1}%
\begin{enumerate}
    \item[1)] How do you think the text-to-image generation system works? [open-ended]
    \item[2)] What is the potential of these systems and how do you think they will be applied in the future? [open-ended]
    \item[3)] What do you think are the dangers of text-to-image generation for you personally? And what are the dangers for society? [open-ended]
\end{enumerate}%
\subsection{In response to the given scenario of Jason submitting an AI-generated artwork to an art contest:}%
\label{appendix:A2}%
\begin{enumerate}%
    \item[4)] Please rate the ethicality of Jason's behavior, if Jason clearly labelled the image as ``generated by AI''. [Likert scale from 1 -- Not At All Ethical to 7 -- Highly Ethical]
    \item[5)] Please rate the ethicality of Jason's behavior, if Jason did not label the image as ``generated by AI''. [Likert scale from 1 -- Not At All Ethical to 7 -- Highly Ethical]
    \item[6)] Please rate the ethicality of Jason's behavior, if the image was generated from a prompt NOT written by Jason. Likert scale from 1 -- Not At All Ethical to 7 -- Highly Ethical]
    \item[7)] Please briefly justify your choices from above. [open-ended]
    \item[8)] Should Jason be required to disclose his text prompt when entering the art competition? [multiple choice: Yes/No/Not sure]
\end{enumerate}
\subsection{Experience with text-to-image generation:}%
\label{appendix:A3}%
\begin{enumerate}
    \item[9)] Have you used text-to-image generation before (e.g., DALL-E, Midjourney, Stable Diffusion, Disco Diffusion, or other systems)? (multiple choice: Yes/No)
    \item[10)] Which of the following systems have you used? (multiple choice; Midjourney/DALL-E Mini, Craiyon/DALL-E 2/Stable Diffusion/DISCO Diffusion/VQGAN-CLIP/Other(s); conditional item)
    \item[11)] How often do you use text-to-image generation? (multiple choice: Never/Rarely/Almost daily/At least once a day/Multiple times a day; conditional item)
    \item[12)] Please give an estimate how many prompts for text-to-image generation you have written (open-ended; conditional item)
    \item[13)] Please describe a specific example of generating an image from prompts. What was your general process for coming up with prompts? (open-ended; conditional item)
    \item[14)] What has obstructed you from using text-to-image generation so far? (multiple choice: I have never heard of it before/It is too difficult to learn/It is too technical/I do not have time/I do not have a use case for it/Other reason(s); conditional item)
\end{enumerate}
\subsection{Importance of text-to-image generation:}%
\label{appendix:A4}%
\begin{enumerate}
    \item[15a)] Professionally, how important is text-to-image generation to you currently? (multiple choice: 1 -- Not At All Important to 7 -- Extremely Important)
    \item[15b)] Professionally, how important is text-to-image generation to you in the future? (multiple choice: 1 -- Not At All Important to 7 -- Extremely Important)
\end{enumerate}
\subsection{Demographics:}%
\label{appendix:A5}%
\begin{enumerate}
    \item[16)] What is your age? [open-ended]
    \item[17)] What is the highest educational degree you have obtained? [multiple choice: No degree/Professional training/Some college-level courses/Bachelor/Master of Arts/Master of Science/Doctoral degree]
    \item[18)] What is you educational background (e.g., Biology, Computer Science, Mathematics, Art History, etc.) [open-ended]
    \item[19)] Are you currently a student? [multiple choice: Yes/No]
    \item[20)] Do you consider yourself an artist? [multiple choice: Yes/No]
    \item[21)] What art do you create? (conditional item) [open-ended]
\end{enumerate}

}%

\end{document}